\documentclass[aps,pre,superscriptaddress,twocolumn,showpacs]{revtex4}
\usepackage{bm}
\usepackage{amsmath}
\usepackage{amsfonts}%
\usepackage{amssymb}%
\usepackage{graphicx}

\newcommand{\kt}{k_B T}
\newcommand{\var}{\mathrm{Var}}
\newcommand{\std}{\mathrm{Std}}

\newcommand{\la}{\langle}
\newcommand{\ra}{\rangle}

\begin{document}
\title{Fluctuating hydrodynamic modelling of fluids at the nanoscale}
\author{G. De Fabritiis}
\email[]{gdefabritiis@imim.es}
\affiliation{Computational Biochemistry and Biophysics Lab (GRIB/IMIM-UPF), Barcelona Biomedical Research Park (PRBB), 
C/ Dr. Aiguader 88, 08003, Barcelona, Spain}
\author{M. Serrano}
\email[]{mserrano@fisfun.uned.es}
\affiliation{Depto.  F\'{\i}sica Fundamental, Facultad de Ciencias, UNED, Paseo Senda del Rey 9, 28040 Madrid, Spain.}
\author{R. Delgado-Buscalioni}
\email[]{rafa@ccia.uned.es}
\affiliation{Depto. Ciencias y T\'ecnicas Fisicoqu\'{\i}micas, 
Facultad de Ciencias, UNED, Paseo Senda del Rey 9,  Madrid 28040, Spain.}
\author{P. V. Coveney}
\email[]{p.v.coveney@ucl.ac.uk}
\affiliation{Centre for Computational Science, Department of Chemistry, University College
London, 20 Gordon Street, WC1H 0AJ London, U.K.}

\begin{abstract}
  A good representation of mesoscopic fluids is required to combine with  
 molecular simulations at larger length and time scales 
(De Fabritiis {\it  et. al}, Phys. Rev. Lett. 97, 134501 (2006)).  However,
accurate computational models of the hydrodynamics  of nanoscale molecular
assemblies are lacking, at least in part  because of the stochastic character
of the underlying fluctuating hydrodynamic equations.  Here we derive a
finite volume discretization of the compressible isothermal fluctuating
hydrodynamic equations  over a regular grid in the Eulerian reference system.
We apply it to  fluids such as argon at arbitrary densities and water under
ambient conditions.  To that end, molecular dynamics simulations are used to
derive the required fluid properties. The equilibrium state of the model is
shown to be thermodynamically consistent and  correctly reproduces linear
hydrodynamics including relaxation of sound and shear modes. We also consider
non-equilibrium states involving diffusion and convection in cavities with
no-slip boundary conditions. 
\end{abstract}
\maketitle

\section{Introduction}

>From a continuum perspective the fundamental equations underlying
hydrodynamics at the mesoscale are the well known fluctuating hydrodynamics
(FH) equations \cite{landau59}. The FH equations are stochastic partial
differential equations that reduce to the Navier-Stokes equations in the limit
of large volumes.  In fact, at scales of nano to micro meters, thermal
fluctuations cannot be neglected, but must be incorporated as random terms in
the momentum and energy equations of hydrodynamics because they are responsible
for the mechanical and thermal energy processes underlying Brownian motion.
The hydrodynamics of nanoscopic quantities of liquids, in particular
for water models used in molecular dynamics simulations such as TIP3P
\cite{Jorgensen}, is relevant in many biological and technological
applications.

Novel multiscale modelling techniques via domain decomposition
(particle-continuum hybrid approach) require a very accurate description of
thermodynamics and hydrodynamics at the mesoscale level \cite{busH1,
defabritiis06hmd}. In these methods a large part of the system is resolved with
a continuum model (CFD) and a smaller part using full-atom molecular dynamics
(MD).  An exact match between the local thermodynamic and hydrodynamic
properties of the continuum and the molecular system is required for such a
scheme to work properly.  This matching enables a seamless coupling such that
the behaviour  of molecular and hybrid simulations are indistinguishable
\cite{defabritiis06hmd}.  Depending on the process or regime being considered
fluctuations may play an important r\^ole \cite{busF}.  For these reasons, the
fluctuating hydrodynamics model described here has been used in the first
hybrid MD-FH model for water which includes mass and momentum fluctuations as
well as for propagation of sound waves across an hybrid interface
\cite{defabritiis06hmd,coveney06coupled} because of the need to have 
a fine control over the characteristics of the fluid to much the molecular description 
(shear and bulk viscosities, equation of state and thermodynamic fluctuations).

An accurate code for fluctuating hydrodynamics, possibly interfaced with
molecular dynamics, would be a useful tool for nanoscale computational fluid
dynamics (CFD) simulation, including {\it inter alia} microfluidic devices
\cite{atencia05}.  These devices are essentially hydraulic micro machines which
are able to process nano liters of reagents. These volumes are too large to be
simulated by molecular dynamics, while, on the other hand, a standard CFD code
cannot handle fluctuations at all. 

A general purpose FH solver could also be used to provide an implicit
hydrodynamic solvent for solute particles (polymers, colloids, etc.). Solvent
molecules often comprise the computationally most expensive part of any
molecular or coarse-grained simulation, but in some instances the solvent could
be approximated by an ``implicit description'', retaining only the hydrodynamic
contribution to the solute dynamics. A possible approach was first illustrated
in \cite{dunweg99}.  A solver of FH for the hydrodynamic description of these
hybrid models could be employed  to study the effect on polymer collapse of changing
solvent characteristics \cite{defabritiis06his} by  tuning the characteristics of the fluids such
as their viscosities.

In recent years, many computational models have been devised which provide a
discrete representation of FH \cite{espanol03review,succi01,malevanets99}, e.
g.  dissipative particle dynamics (DPD) \cite{hoog92,espanol95} and the lattice
Boltzmann method (LBM) \cite{higuera89,benzi92,succi01,succi01multiscale,karlin06} later extended to include
thermodynamic fluctuations \cite{Ladd05,ladd94,adhikari05}. 
Following a continuum approach,  the fundamental equations of fluctuating
hydrodynamics can also be resolved directly via finite differences or finite
volume schemes \cite{garcia87,serrano01}.  Most of the previous schemes
proposed for solving fluctuating hydrodynamics considered only the gas phase
\cite{garcia87,garcia87prl} and have focused on the linear regime where a
closed set of equations for the fluctuating quantities or for their mutual
spatial correlations can be derived \cite{garcia87prl}. The main difficulty in
devising a discretization of the FH equations is that the precise form of the
required fluctuation-dissipation relations depends on the discretization scheme
and, in general, it does not coincide with the fluctuation-dissipation
relations of the continuum description. This statement applies for Lagrangian
FH models \cite{generic2,flekkoy00,serrano01}, for fluctuating lattice
Boltzmann models \cite{adhikari05,Ladd05,ladd94} and  for the Eulerian FH
description.  Moreover the resulting equations are stochastic in nature which
adds  extra complications to the integration methods
\cite{kloeden92,defabritiis06_trotter, serrano06_trotter}. 

 For these reasons, a simple and easy-to-implement general purpose solver of
the compressible FH equations allowing fluid specificity and non-linear
hydrodynamic coupling is not readily available. It is the purpose of this paper
to address this lacuna. We present a finite volume discretization of the
compressible isothermal fluctuating hydrodynamic equations, based on an
Eulerian description on a regular grid. The model provides a thermodynamically
consistent coarse-grained representation of nano liter portions of real fluids
ranging from gases (argon) to liquids (water).

The outline of the paper is the following: The fluctuating hydrodynamic
Eulerian solver is described in section \ref{FVDFH}.  In section
\ref{accuracy} we study the numerical accuracy of the scheme by comparing the
input values of the viscosities with the effective ones measured from the
hydrodynamic solver based on the relaxation of sound and shear waves.  In
section \ref{hydro_equ} we assess the validity of the description of the
equilibrium state, showing that fluctuations are correctly generated,
propagated and dissipated. To that end we calculate the time correlation
functions of the different fluctuating variables of one fluid cell (density
and velocity) and compare them with the corresponding grand canonical result.
In section \ref{non-eq} we consider non-equilibrium states in closed systems
with rigid walls, using the no-slip boundary condition: we test them for
Couette, Poiseuille and cavity flows. 
Finally, we summarise our findings in section \ref{summary}.

\section{A finite volume discretization of fluctuating hydrodynamics}
\label{FVDFH}

Our proposed mesoscopic model is a finite volume discretization of FH
\cite{flekkoy00,serrano01} over a regular lattice in the Eulerian frame of
reference. In this case, we concentrate on the description of an isothermal
compressible fluid. This sort of description can be generalized
straightforwardly to non-isothermal states for fluids with vanishingly small
thermal expansion, for which the energy equation decouples from the mass and
momentum equations \cite{garcia87}.  Extensions to include energy flows will
be considered elsewhere. We thus require the equations of fluctuating
hydrodynamics describing the conservation of mass and momentum,
\begin{align}
  \partial_{t}\rho & =-\partial_{\beta}g_{\beta},\nonumber \\
  \partial_{t}g_{\alpha} & = -\partial_{\beta} \left( g_{\beta} v_{\alpha}+
\Pi_{\alpha\beta}+\widetilde{\Pi}_{\alpha\beta} \right)  ,\label{fheq}%
\end{align}
where $\rho({\bf r},t)$ is the density field of the fluid, $v_{\alpha}({\bf
  r},t)$ is the continuous velocity field in the component $\alpha$,
$g_{\beta}({\bf r},t)=\rho ({\bf r},t)v_{\beta}({\bf r},t)$ is the momentum
field and we have used the repeated suffix convention for summation over
repeated indices.  $\Pi_{\alpha\beta}({\bf r},t)$ and
$\widetilde{\Pi}_{\alpha\beta}({\bf r},t)$ are respectively the average
(Navier-Stokes) and fluctuating stress tensor fields.  The average stress
tensor is defined as $\mathbf{\Pi}=(p+\pi)\mathbf{1}+\overline{\mathbf{\Pi}}$,
where $p$ is the thermodynamic pressure given by the equation of state for the
fluid, $\pi = - \zeta \partial_{\gamma} v_{\gamma}$ and
$\overline{\Pi}_{\alpha\beta} =- \eta \left( \partial_{\alpha}v_{\beta} +
  \partial_{\beta}v_{\alpha} - 2 D^{-1}\partial_{\gamma}v_{\gamma}
  \delta_{\alpha\beta} \right)$ where $\eta$ and $\zeta$ are the shear and
bulk viscosity respectively and $D$ is the spatial dimensionality.

The equation of state $p=p(\rho,T)$ for Lennard-Jones (LJ) fluids (like argon)
has been studied by several authors (see e.g., Ref. \cite{Johnson93}), as well
as the transport coefficients of the LJ fluid \cite{Heyes88, Borgelt90}.  By
contrast, the equation of state of the TIP3P model for water \cite{Jorgensen}
(chosen  due to its importance in biological applications) has received less
attention \cite{Jeffrey}.  In Appendix \ref{Argon_water} we provide a
parametric study of the equations of state of the fluids considered here
(argon and TIP3P water model), performed via molecular dynamics (MD)
simulations.  From this study we obtain a second order polynomial fit for
$p=p(\rho,T)$ which provides the equation of state of our FH model.  This
procedure is required, for instance, to provide the level of accuracy for the
thermodynamic pressure within hybrid MD simulations \cite{defabritiis06hmd}.
We have also calculated 
the transport coefficients of water via non-equilibrium molecular dynamics,
which are in agreement with those reported in the literature \cite{Guo01}.

The fluctuating stress tensor $\widetilde{\Pi}_{\alpha\beta}$ (see
Ref.\cite{landau59}) is a random Gaussian matrix with zero mean and
correlations given by
\begin{align}
 \langle\widetilde{\Pi}_{\alpha\beta}(\mathbf{r}_1,t_1)
\widetilde{\Pi}_{\delta\gamma}(\mathbf{r}_2,t_2)\rangle =& 2 k_B T C_{\alpha
  \beta \gamma \delta} \delta(t_1-t_2) \delta(\mathbf{r}_1-\mathbf{r}_2),
\end{align}
where $C_{\alpha \beta \gamma \delta}=\left[ \eta
  (\delta_{\alpha\delta}\delta_{\beta\gamma}
  +\delta_{\alpha\gamma}\delta_{\beta\delta}+ (\zeta-\frac{2}{D}\eta)
  \delta_{\alpha\beta}\delta_{\delta\gamma} \right]$, $k_B$ is the Boltzmann
constant and $T$ is the temperature. Note that this spatial delta-correlated
quantity, in the discrete limit of a small volume and small time interval, can be
rewritten as
\begin{align}
  \langle\widetilde{\Pi}_{\alpha\beta}(\mathbf{r}_1,t_1)
  \widetilde{\Pi}_{\delta\gamma}(\mathbf{r}_2,t_2)\rangle \approx& \frac{2 k_B
    T}{\Delta t \Delta V}C_{\alpha \beta \gamma \delta},
\end{align}
where $\Delta V$ is the small volume element of fluid and $\Delta
t$ is the time step.

The fluctuating hydrodynamic equations (\ref{fheq}) are balance equations of
the form $\partial_{t}\phi({\bf r},t)=- \mbox{\boldmath$\nabla$}\cdot
\mathbf{J}^{\phi}$ for mass and momentum which can be integrated by
considering a finite volume discretization.  In what follows we will derive a
finite volume discretization of the equations of fluctuating hydrodynamics in
the Eulerian system of reference.  We first partition the space into $N$ space
filling volumes $V_{k}$ (in our case a regular Cartesian lattice is used) with
$k=1,...,N$ to integrate Eqs. (\ref{fheq}) over the volume $V_{k}$ and apply
Gauss's theorem
\begin{equation}
\frac{d}{dt}\int_{V_{k}}\phi({\bf r},t)d {\bf r}=
\sum_{l} \mathbf{J}^{\phi}_{kl}\cdot\mathbf{e}_{kl} A_{kl},
\end{equation}
where $\mathbf{e}_{kl}$ is the unit vector perpendicular to the contact
surface of area $A_{kl}$ from volume $l$ to volume $k$. The summation is over
all the $l$ volumes that are in contact with fluid volume $k$.  By defining
$M_{k}^{t}=\int_{V_{k}}\rho({\bf r},t)d {\bf r}$ as the mass inside a generic
volume $V_k$, and $\mathbf{P}_{k}^{t}= \int_{V_{k}}\rho({\bf r},t) {\bf
  v}({\bf r},t)d {\bf r}$ its momentum, we thus build dynamical equations
corresponding to discrete extensive variables which replace Eqs.(\ref{fheq})
governing the time evolution of the intensive continuum fields. These new
equations are given by
\begin{align}
dM_{k}^{t}  & =%
\sum_l \mathbf{g}_{kl} \cdot\mathbf{e}_{kl}A_{kl}dt,\nonumber\\
d\mathbf{P}_{k}^{t} & =\sum_l \left(\mathbf{v}_{kl} \ \mathbf{g}_{kl} 
+\mathbf{\Pi}_{kl}
\right)\cdot
\mathbf{e}_{kl}A_{kl}dt
+d\widetilde{\mathbf{P}}_{k}^{t},\label{mass_momupdate}
\end{align}
where we have approximated the mass flux ${\bf J}^\rho_{kl}$ with
$\mathbf{g}_{kl}=\frac{1}{2}(\rho_{k}+\rho_{l})\frac{1}{2}(\mathbf{v}_{k} +
\mathbf{v}_{l})$, the velocity on the surface $kl$ as
$\mathbf{v}_{kl}=\frac{1}{2}(\mathbf{v}_{k} + \mathbf{v}_{l})$, the average
stress tensor on the surface as $\mathbf{\Pi}_{kl}=\frac{1}{2}[
(p_{l}+\pi_{l})\mathbf{1}+\overline{\mathbf{\Pi}}_{l}]$ and
$d\widetilde{\mathbf{P}}_{k}^{t}$ indicates the momentum change due to the
fluctuating part of the pressure tensor, all at time $t$.
  
An essential component for the discretization of a stochastic mesoscopic model
is the balancing of dissipative and fluctuating components, otherwise the
fluctuation-dissipation theorem would not be satisfied.  There are at least
two ways to satisfy this condition, either by using the Fokker-Planck
equations mathematically equivalent to the stochastic differential equations
(SDE \ref{mass_momupdate}) and equating dissipative and diffusive terms
weighted over the Gibbs ensemble distribution \cite{flekkoy00} or by using the
GENERIC formalism \cite{generic0,serrano01}.  By choosing the gradient
discretization provided in \cite{serrano01} a lot of long algebra is avoided;
we find discrete versions for
\begin{align}
\overline{\Pi}_{k}^{\alpha\beta}  & =\frac{\eta_{k}}{V_{k}}\sum_{l}\left[
\frac{A_{kl}}{2}(e_{kl}^{\alpha}v_{l}^{\beta}+e_{kl}^{\beta}v_{l}^{\alpha
})-\frac{\delta^{\alpha\beta}}{D}A_{kl}e_{kl}^{\gamma}v_{l}^{\gamma}\right]
, \nonumber\\
\pi_{k}  & =\frac{\zeta_{k}}{V_{k}}\sum_{l}\frac{A_{kl}}{2}e_{kl}^{\beta}%
v_{l}^{\beta},\label{presstens}
\end{align}
and the fluctuating component of the momentum equation given by
\begin{eqnarray}
d\widetilde{\mathbf{P}}_{k}^{t}&=&\sum_{l}\frac{A_{kl}}{2}  
\sqrt{4k_BT_{l}\frac{\eta_{l}}{V_{l}}}d\overline{{\bf W}}_{l}^{S} 
\cdot\mathbf{e}_{kl} \nonumber\\ 
&+&\sum_l \frac{A_{kl}}{2} \sqrt{  2Dk_BT_{l}\frac{\zeta_{l}}{V_{l}}} 
\frac{tr[d\mathbf{W}_{l}]}{D} \mathbf{e}_{kl},
\end{eqnarray}
where $d{\bf W}_l$ is a $D\times D$ matrix ($D=3$ in three dimensions) of
independent Wiener increments satisfying $\langle d{\bf
  W}_k^{\alpha\beta}d{\bf W}_l^{\gamma\delta}\rangle=\delta_{kl}
\delta_{\alpha\gamma}\delta_{\beta\delta} dt$ and $d\overline{{\bf
    W}}_{l}^{S}$ is a traceless symmetric random matrix defined as
\begin{eqnarray}
d\overline{{\bf W}}_{l}^{S}=\frac{(d{\bf W}_l+d{\bf W}_l^T)}{2}
-\frac{tr[d{\bf W}_l]}{D}{\bf  1}.
\end{eqnarray}
The resulting set of stochastic differential equations is integrated using a
simple stochastic Euler scheme in the present work. Note, however, that other
more accurate stochastic integration schemes for mesoscopic models have
recently been proposed based on the Trotter expansion in the stochastic case
\cite{defabritiis06_trotter,serrano06_trotter}.
Improvements to the solver for  the spatial regular grid and for the time integration 
scheme have not been considered for the present scheme because at these scales the Reynolds number
is usually low and the computational limitation comes rather from the molecular dynamics component. 
However, generalizations  to unstructured grids are straightforward. In particular, in a hybrid MD-CFD model
finer cells should be located near the MD region and coarser, bigger cells further away from the MD domain.

\section{Accuracy of the scheme} 
\label{accuracy}
To assess the accuracy of the numerical scheme, we measure the effective
viscosities and sound velocity computed from the hydrodynamic solver and
compare them with the input values.  The transport coefficients are measured
from the relaxation of transversal and longitudinal waves in the deterministic
limit. We give more details in Appendix \ref{appendix1}.  In the following
tests we consider argon at temperature $T=300$ K and mass equilibrium density
$\rho_e=0.6$ g/mol/\AA$^{3}$ and TIP3P water. The corresponding values of
the dynamic shear and bulk viscosity for argon and water are shown in table
\ref{argon_transport_coeff}, along with the isothermal sound velocity, $c_T^2
\equiv (\partial P/\partial \rho)_T$ and sound absorption coefficient
$\Gamma_T$.  We consider a periodic 3D cubic domain of length $L$; thus, the
permitted wavelengths are $k_n=2\pi n/L$, and we excite the longest wavelength
of the system, of wave-number $2\pi/L$.
\begin{table}[h]
\caption{\label{tab:eosargon} Some properties of 
argon and water at $T=300$ K at  the mass densities $\rho$
considered; $m$ is the molecular mass. Note that length, time and mass units are
\AA, ps and g/mol respectively.  The properties
displayed are shear viscosity ($\eta$), bulk viscosity ($\zeta$), isothermal
sound speed ($c_T$) and isothermal sound absorption ($\Gamma_T$) in
corresponding units.}
\begin{tabular}{|c|c|c|c|c|c|c|}
\hline
liquid & $m$ &$\rho$ & $\eta$ & $\zeta$ & $c_T$&  $\Gamma_T$\\
\hline
argon & 39.948 &0.60  & 5.474 & 1.823 & 5.614 &  7.612\\
water & 18.015 &0.632 & 53.71 & 127.05 & 14.75 & 157.17\\
\hline
\end{tabular}
\label{argon_transport_coeff}
\end{table}

\subsection{Transversal wave}
The wave-vector of a transversal wave is perpendicular to its velocity, ${\bf
  k} \bot {\bf v}$.  Consider an initial perturbative velocity ${\bf
  v}=(v_0\sin(k z),0,0)$, with ${\bm k}=(0,0,k)$.  For small perturbations
around equilibrium, the linearized solution for the momentum density field in
Fourier space given in Eqs.(\ref{dens}) is
\begin{eqnarray}
{\bf g}({\bm k},t)&=& \exp\{-\nu k^2 t\} {\bf g}(\bm k,0),
\end{eqnarray}
while in real space, the time dependent velocity field reads
\begin{equation}
v_x(t)=v_0 \sin(k z)\exp\{-\nu k^2 t\};
\label{theory_trans}
\end{equation}
in addition  $v_y(t)=v_z(t)=0$ and $\rho(t)=\rho_e$.
\vspace{0.5cm}
\begin{figure}[tb!]
\begin{center}
\includegraphics[width=6cm,angle=0]{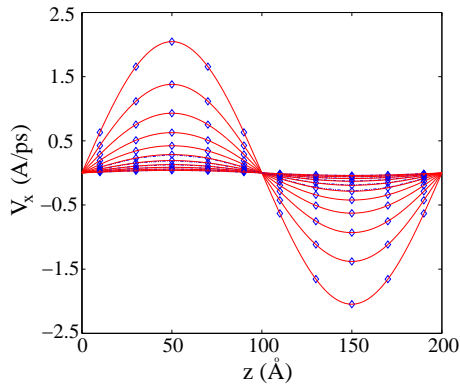}
\caption[]{
  Velocity
  field $v_x$ as a function of $z$ (in \AA) for this decaying
  transversal wave.  Diamonds correspond to simulation results and the
  continuum lines correspond to the theoretical profile at snapshots
  corresponding to times $t=0,50,100,\cdots,750$ $ps$. 
At time $t=0$ $ps$ the amplitude is maximum while
at $t=750$ $ps$ the amplitud is mininum. }
\label{transversalfourier}
\end{center}
\end{figure}

In Fig.\ref{transversalfourier} we plot some snapshots of the velocity field
for a deterministic simulation. This case represent a three dimensional
simulation box of $200 \times 200 \times 200$  \AA$^3$ ($10 \times 10 \times 10$
cells) which corresponds to a spatial resolution $\delta=20$ \AA. The applied
initial velocity amplitude is $v_0=2.04$ \AA/ps, the argon input viscosity at
$300$ K is $\eta=5.4744$ and the mean mass density is $\rho=0.6$ g/mol/\AA$^3$.
The theoretical agreement with expression (\ref{theory_trans}) is remarkable.

\begin{figure}[tb]
\begin{center}
  \includegraphics[width=7.5cm,angle=0]{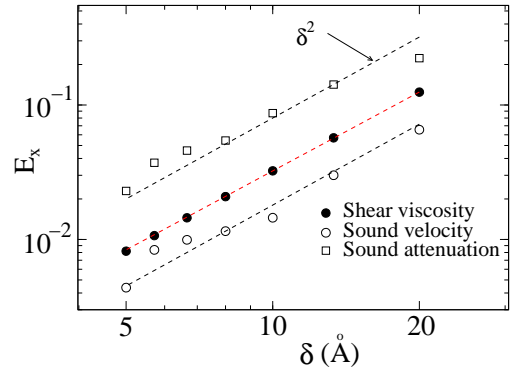}
\caption[]{Relative error for the shear viscosity, sound absorption
coefficient and sound velocity as a function of the spatial resolution
$\delta$.  For any fluid parameter (e.g. $\nu$) the relative error is
defined as $E_{\nu}\equiv |\nu_{num}-\nu|/\nu$, where $\nu$ is the
input value and $\nu_{num}$ that measured from the relaxation of the
corresponding hydrodynamic mode. }
\label{resolution}
\end{center}
\end{figure}

For a closer inspection of the numerical accuracy of the scheme we compared
the effective (or numerical) shear viscosity $\nu_{num}$ with the input value
$\nu$. The value of $\nu_{num}$ was measured by fitting the decay of the
Fourier component of the transversal momentum to a simple exponential
function.  The relative error in viscosity $E_{\nu}\equiv |\nu_{num}-\nu|/\nu$
is shown in Fig.\ref{resolution} against the spatial resolution $\delta$,
given by the distance between contiguous cells.  The trend obtained is
$E_{\nu} \propto \delta^{1.94}$, showing that our spatial discretization
method is of second order. As an example, a continuum cell size of $\delta=20$
{\AA} will give an error in the viscosity around $12\%$ while $\delta=15$ {\AA}
will reduce the relative error to $5\%$ for the pertrubation applied here.

\subsection{Longitudinal wave}
If the equilibrium state of the fluid is initially perturbed with a momentum
field ${\bf g}_0=(g_0\sin k x,0,0)$, a longitudinal sound wave is created with
a wave-vector parallel to the momentum perturbation ${\bm k}= (2\pi/L,0,0)$. As
shown in Appendix \ref{appendix1}, two travelling sound modes propagate at the
sound velocity $c_T$, creating a standing wave in the periodic domain.
According to Eqs. (\ref{dens}) the Fourier components of the deviation from
the equilibrium state for the density and velocity of the sound modes are
given by
\begin{eqnarray}
\delta \rho({\bm k},t)&\propto&-i \exp\{-\Gamma_Tk^2 t\}
\sin (c_T k  t) \frac{{\bm k}\cdot {\bf g}_0}{c_T},\nonumber\\
\delta {\bf g}({\bm k},t)&\propto& \exp\{-\Gamma_Tk^2 t\}
\cos(c_Tk t) {\bm k} \frac{{\bm k}\cdot {\bf g}_0}{c_T},
\label{theory_long}
\end{eqnarray}
while in real space the density and velocity fields evolve like
\begin{eqnarray}
\rho(t)&=&\rho_e+\frac{g_0}{c_T}\cos(kx)\sin(c_T k t)\exp\{-\Gamma_T k^2 t\},\nonumber\\
v_x(t)&=&\frac{g_0}{\rho_e}\sin(kx)\cos(c_T k t)\exp\{-\Gamma_T k^2 t\}
\label{theory_long_real}
\end{eqnarray}
and the other velocity field components remain at rest. Note that $\rho_e$
denotes the equilibrium density.

\begin{figure}[tb!]
\begin{center}
\includegraphics[width=6cm,angle=0]{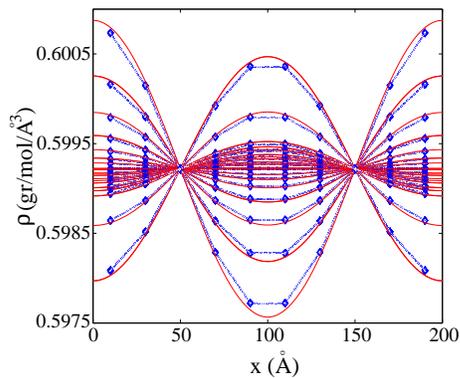}
\caption[]{
  The mass density $\rho$ as a function of $x$ (in \AA) for a decaying
  longitudinal wave.  Diamonds correspond to simulation results and the
  continuum lines correspond to the theoretical profile given by Eq.
  (\ref{theory_long_real}). The snapshots correspond to the times $t=0,25, 50,
  \cdots,750$ $ps$. The agreement is remarkable. At time $t=0$ $ps$ the
  amplitude is maximum while at $t=750$ $ps$ the amplitud is mininum.
}
\label{longitudinalfourier_deterministic}
\end{center}
\end{figure}
In Fig.\ref{longitudinalfourier_deterministic} we plot the density field as a
function of $x$ for a deterministic three dimensional simulation of size $200
\times 200 \times 200$ \AA$^3$ ($10 \times 10 \times 10$ cells). In order to
keep the system in the linear regime, the applied initial velocity amplitude is
set to $v_0=0.0204$ \AA/ps and the argon input mean mass density is $\rho=0.6$
g/mol/\AA$^3$. The best fit to the absorption coefficient and sound velocity
are $\Gamma_T=7.24$ and $c_T=5.42$, providing relative errors of $4.8\%$ and
$3.6\%$ respectively. Figure \ref{resolution} also shows the relative error in
the sound absorption coefficient and sound velocity versus the spatial
resolution.  Note that the relative error in both quantities also decays
roughly like $\delta^{2}$ which corroborates the second order spatial
resolution of the scheme again.

\section{The equilibrium state}
\label{hydro_equ}

Within a fluid volume, stress fluctuations arise due to forces involved in the
random sequence of molecular collisions.  This fluctuating force generates
momentum fluctuations.
The amplitudes of mass and momentum fluctuations are determined by
thermodynamic constraints. Each fluid cell is an open system with constant
volume and temperature, in other words, belonging to the grand canonical
ensemble. The variance of momentum and mass of a cell ``c'' are
\cite{landau59}
\begin{eqnarray}
\label{momfluc}\var\left[\bm P_c\cdot {\bm e}\right] &=& \rho V_c \kt, \nonumber\\
\label{densfluc} \var\left[M_c\right]&=& \frac{\rho V_c \kt}{c_T^2},
\end{eqnarray}
where ${\bm e}$ is a unit surface vector, the mean mass of the cell is $\rho
V_c$ and $c_T$ is the isothermal sound velocity. On the other hand, these
spontaneous mass and momentum fluctuations are transported through the fluid
and dissipated following the same mechanism underlying the hydrodynamic modes
explained in Appendix \ref{appendix1}, i.e., either via shear or sound
modes.

\subsection{Amplitude of fluctuations}
\begin{figure}[tb!]
\begin{center}
\includegraphics[width=7.5cm]{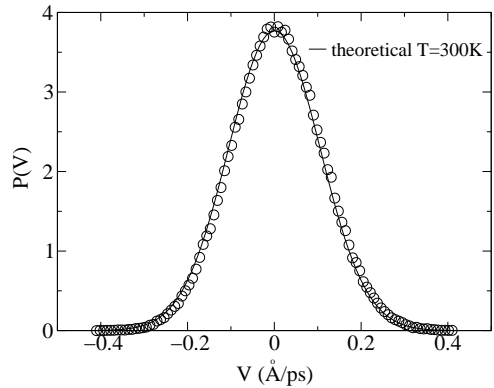}
\caption[]{The equilibrium
  distribution of the $x$ component of the velocity at one fluid cell of volume
  $37.5$ {nm}$^3$ in a simulation of argon (circles) at $T=300$ K and
  $\rho=0.6$ g/mol/\AA$^3$ compared with the theoretical normal distribution
  (continuum line).  In this simulation $\Delta t= 20$ fs and the best normal
  fit to the numerical distribution yields $T_{num}=296.28$ K, that is a
  relative error of around 1.2\%.}
\label{vx_vy_vz_eq}
\end{center}
\end{figure}

In this section we consider the equilibrium state of argon at different
densities (from gas to liquid) and water (TIP3P model, see Appendix
\ref{Argon_water}) in order to illustrate that the fluctuations are generated,
transported and dissipated in a thermodynamically and hydrodynamically
consistent way.  First, we confirm that the amplitudes of mass and momentum
fluctuations are consistent with thermodynamic relations \cite{landau59}.  In
the numerical scheme the amplitude of fluctuations is determined, by
construction, via the fluctuation-dissipation theorem \cite{keizer87}.  Figure
\ref{vx_vy_vz_eq} presents a typical distribution of one velocity component
and compares it with the theoretical Maxwellian distribution.  As usual, a
temperature can be extracted from the variance of the velocity distribution.
This ``numerical'' temperature will be labelled as $T_{num}=\sum_{\alpha}
\var[v_{\alpha}]/(3 k_B)$.

\begin{figure}[tb!]
\begin{center}
\includegraphics[width=6cm,angle=0]{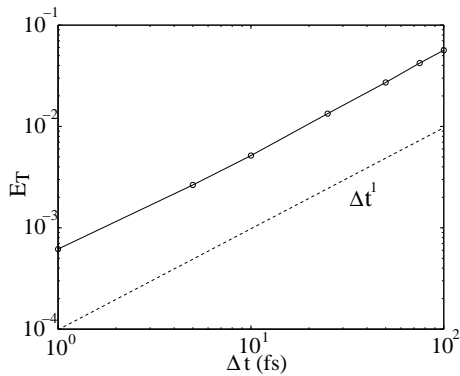}
\caption{Relative error in the temperature of the
  scheme, $E_T=|T_{num}-T|/T$,  with $T_{num}\equiv \sum_{\alpha}
  \var[v_{\alpha}]/(3 k_B)$ and $T=300$ K the input equilibrium temperature.
  Results correspond to three-dimensional simulations of water at $\rho=0.6$ g/mol/\AA$^3$ with
  spatial resolution $\delta = 20$ {\AA} in each direction. }
\label{temp_dt}
\end{center}
\end{figure}

  The accuracy of the stochastic time integrator for the Langevin equation affects the 
  value of the numerical temperature. Figure \ref{temp_dt} shows the
  dependence of the relative error in the mean temperature (defined as
  $E_T=|T_{num}-T_e|/T_e$) with the time step $\Delta t$. Good agreement is
  found and the relative error remains smaller than $10\%$ for $\Delta t
  \leq 100$ fs.

\begin{figure}[tb!]
\begin{center}
\includegraphics[height=6cm,angle=0]{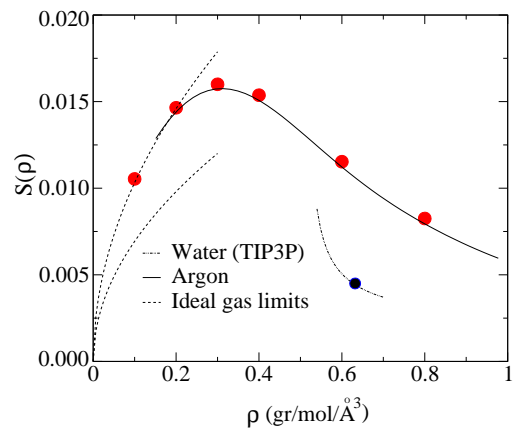} 
\caption{Standard deviation of the mass density $S[\rho]$ in a fluid cell
  with volume $37.5$ {nm}$^3$ of argon and water at $T=300$K. The continuous line
  corresponds to the grand canonical results obtained using the equation of
  state for argon, the dotted lines for water (TIP3P model) $[\rho
  \kt/(c_T^2V_c)]^{1/2}$, and the dashed lines show the ideal gas limits
  $[(m/V)\rho]^{1/2}$ with $m_{Ar}=39.498$ g/mol (upper curve) and
  $m_{H_2O}=18.015$ g/mol (lower curve).  Circles are results from the
  fluctuating hydrodynamics solver (using a water simulation at $\rho=0.632$
  g/mol/\AA$^3$).}
\label{GC}
\end{center}
\end{figure}

Figure \ref{GC} shows the standard deviation of the cell mass density
$\std[\rho_c]$ against the mean density $\rho =M/V$, where $M$ and $V$ are the
total mass and volume of the system.  The grand canonical prediction for the
equilibrium state is $\std[\rho]= [\rho \kt/(c_T^2V_c)]^{1/2}$  and  is 
compared with the numerical simulations.  In the ideal gas limit the sound
velocity is just $\sqrt{\kt/m}$ (with $m$ the molecular mass) and the density
fluctuations increase as $\var[\rho^{(ideal)}]= (m/V)\rho$. As the fluid
becomes denser, it becomes less compressible (the isothermal sound speed $c_T$
increases) and as a consequence in the liquid phase the mass fluctuations
decrease substantially.  Therefore, the largest mass fluctuations are observed
at moderate densities (e.g. around $\rho\simeq 0.3$ g/mol/\AA$^3$ for argon

see Fig.\ref{GC}). Almost perfect agreement is found between
theoretical and numerical results for both argon and water.

\subsection{Correlations at equilibrium}
We now show that fluctuations are transported through the system and
dissipated in a hydrodynamically correct way. This can be shown via the
time correlation of the fluctuating quantities.  As shown in
Ref. \cite{keizer87}, the time correlation of the Fourier components of
mass and momentum satisfy
\begin{eqnarray}
\frac{\la \rho(k,t) \rho (k,0)\ra}{\var[\rho(k,0)]} &=&
\exp\{-\Gamma_T k^2 t\} \cos(c_T k t), \nonumber\\
\frac{\la g_{\|}(k,t)g_{\|}(k,0)\ra}{\var[g_{\|}(k,0)]} 
&=& \exp\{-\Gamma_T k^2 t\}\cos(c_T k t), \nonumber\\ 
\frac{\la g_{\perp}(k,t)g_{\perp}(k,0)\ra}{\var[g_{\perp}(k,0)]} 
&=& \exp\{-\nu k^2 t\}, \nonumber\\ 
\frac{\la \rho(k,t) i\,g_{\|} (k,0)\ra}{\la \rho(k,0) i\, g_{\|}(k,0) \ra } 
&=& \exp\{-\Gamma_T k^2 t\} \sin(c_T k t). 
\label{correlacs}
\end{eqnarray}
In Eqs.(\ref{correlacs}) $g_\|$ indicates the longitudinal momentum, parallel
to the wave vector $\bm k$, while $g_{\perp}$ indicates the transversal
components.  Sound modes couple longitudinal momentum and density, as shown in
Eqs.(\ref{correlacs}).  According to the Landau description of fluctuating
hydrodynamics \cite{landau59}, at a fixed time the stress fluctuations
occurring within different fluid volumes are spatially uncorrelated. This
means that the variance of the Fourier modes of the hydrodynamic variables
($\var[\rho(k,0)]=\la \rho(k,0)^2\ra $) is independent of the wave-number $k$.
Moreover, fluctuations from the equilibrium state are assumed to be small so
that a linear analysis can be applied.  This means that perturbations of
different wavelengths evolve independently as correlations between
fluctuations with different wave-vectors are negligible.

\begin{figure}[tb!]
\begin{center}
\includegraphics[height=7cm,angle=0]{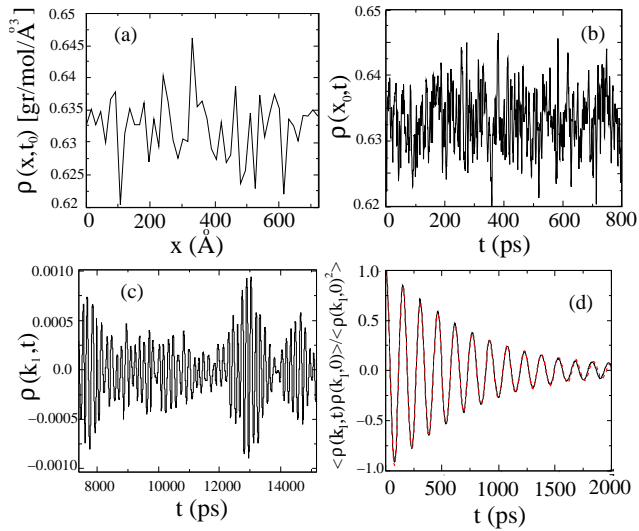}
\caption{Water at $\rho=0.632$ g/mol/\AA$^3$ and $T=300$ K (ambient conditions) 
  within a periodic box of size $50\times 50\times 2250$ \AA$^3$ at equilibrium;
  the FH mesh is comprised of of $1\times 1\times 150$ cells.  (a) The density
  field in the real space, $\rho(x,t_0)$.  (b) The time-dependence of the
  density at one cell $\rho(x_0,t)$.  (c) The Fourier mode $\rho(k_1,t)$
  associated with wavevector $\bm k_1=(0,0,2\pi/L_z)$.  (d) The (normalized)
  time correlation function $\la \rho(k_1,t)\rho(k_1,0)\ra$. Note the
  different time scales associated with variations of quantities in (b),(c)
  and (d). The entire run is of $40$ ns duration.}
\label{figcor1}
\end{center}
\end{figure}

\begin{figure*}[t]
\begin{center}
\includegraphics[width=5cm,angle=0]{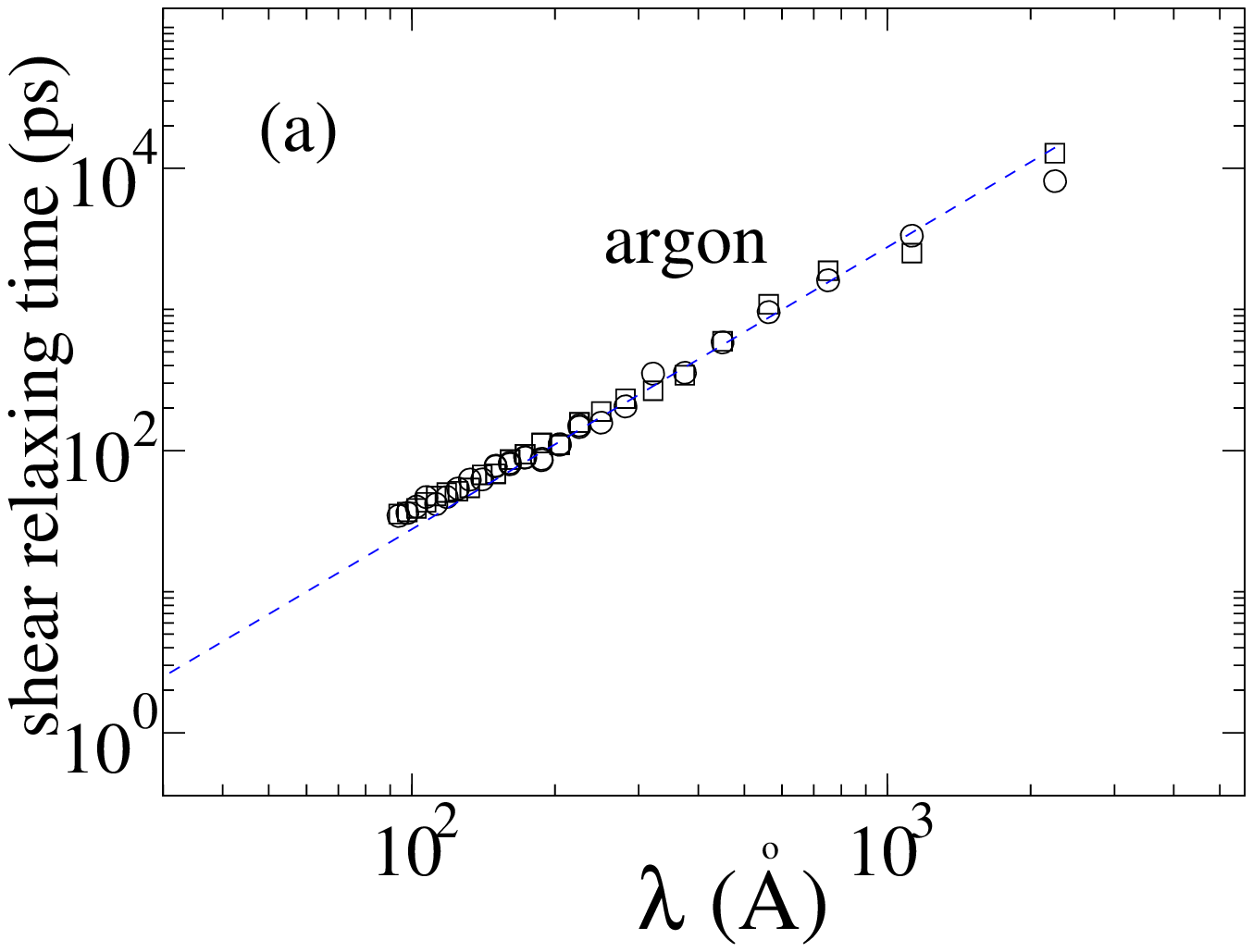}
\includegraphics[width=5cm,angle=0]{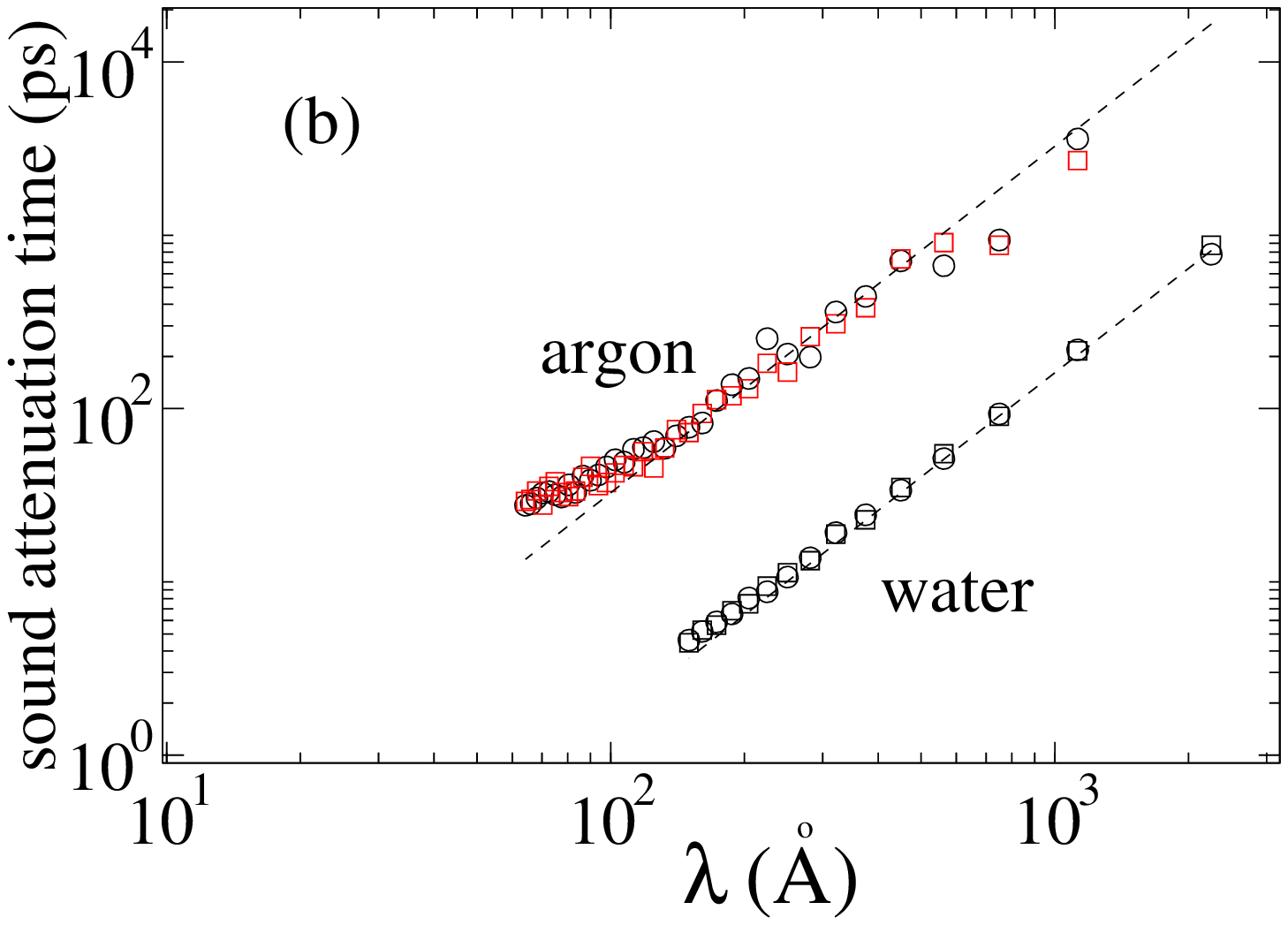}
\includegraphics[width=5cm,angle=0]{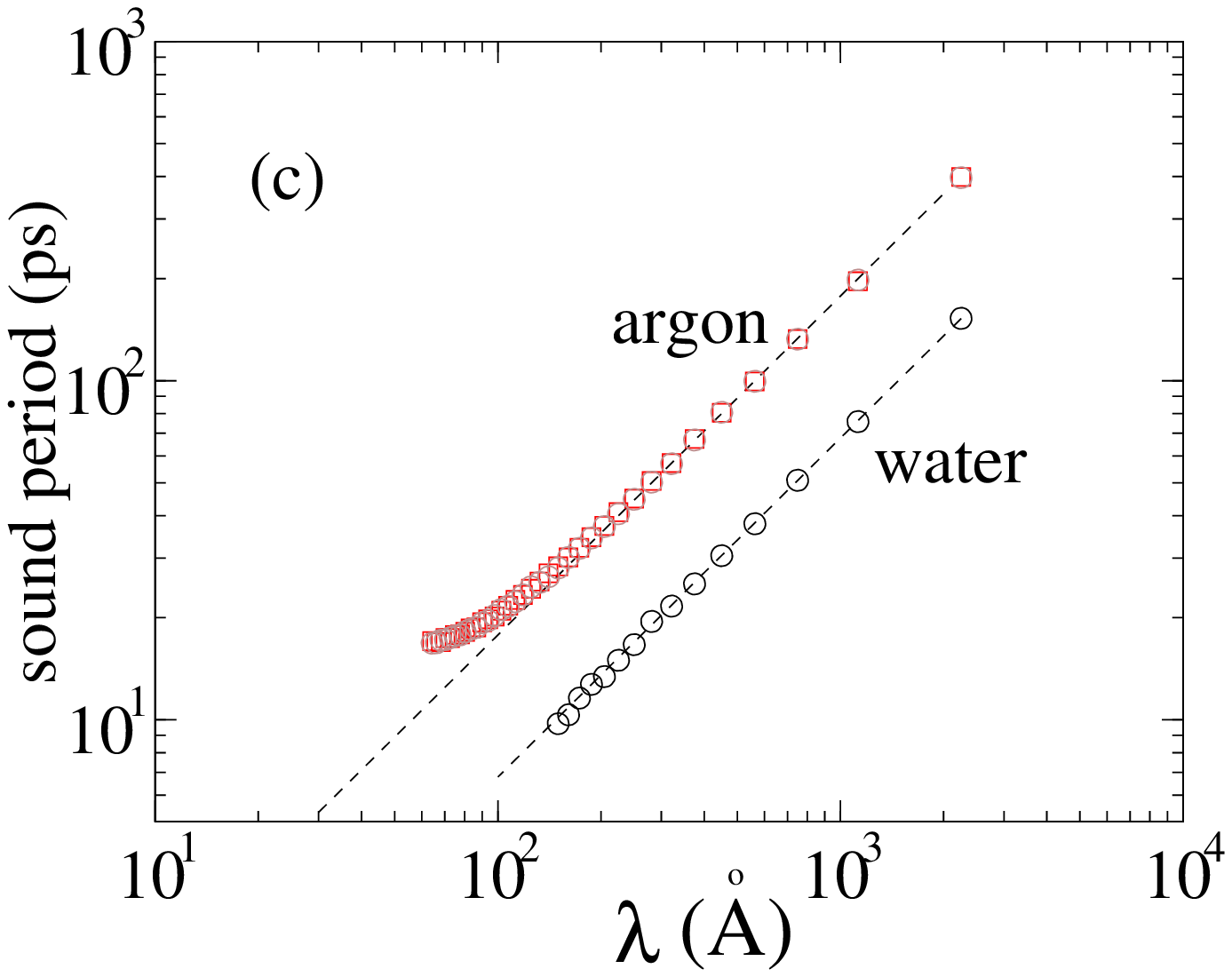}\\
\caption{The best fits to the decay times obtained from the time
  correlations of the Fourier modes of wavelength $\lambda=2\pi/k$.  Symbols
  correspond to numerical results and lines to theoretical predictions (see
  Eqs.  (\ref{correlacs})).  (a) Shear decay times: the lines are the theoretical
  results $(k^2\nu)^{-1}$, (b) sound attenuation time $(k^2 \Gamma_T)^{-1}$ and
  (c) the sound period $\lambda/c_T$. Circles are obtained from the imaginary
  part of $\la h(k,\tau) h(k,0) \ra$ and squares from the real part, where
  $h=v_{\perp}$ (transversal velocity) in (a) while $h=\rho$ in (b) and (c)
  (similar results, not shown,  were obtained for the longitudinal velocity $v_{\|}$).
  Results for water ($c_T=14.75$ \AA/ps, $\nu= 84.98$ \AA$^2$/ps and
  $\zeta/\rho=201.02$ \AA$^2$/ps) correspond to the simulations illustrated in
  Fig.\ref{figcor1}, while argon ($c_T=5.61$ \AA/ps) corresponds to
  $\rho=0.6$ g/mol/\AA$^3$ and $T=300$ K in the same box used in Fig.\ref{figcor1}.}
\label{figcor2}
\end{center}
\end{figure*}

We have evaluated the time correlations of the Fourier components of the
hydrodynamic variables in a periodic domain for the set of wave-numbers
$k_n=2\pi n/L$. These correlations were then fitted to the corresponding
exponentially decaying functions of Eqs.(\ref{correlacs}) to obtain the
effective sound frequency $c_T k$ and the effective decay rates for each
wave-number (i.e. the inverse of $\Gamma_T k^2$ and $\nu k^2$ for sound and
shear, respectively). Calculations were done in a periodic system of size
$50\times 50\times 2250$ \AA$^3$, with a mesh of $1\times 1\times 150$ cells,
and we considered perturbations with wave-vectors ${\bm k}_n=(0,0,2\pi n/L_z)$.
In order to illustrate the mathematical transformations, Fig.\ref{figcor1}
shows the density field in the real space $\rho(x,t_0)$ (a), the time-dependence
of the density at one cell $\rho(x_0,t)$ (b), the Fourier mode $\rho(k_1,t)$
associated with $k_1=2\pi/L_z$ (c) and, finally in (d), the time correlation function
$\la \rho(k_1,t)\rho(k_1,0)\ra$ together with the best fit obtained to the theoretical
exponential decay of the sound mode.  The best fits to the effective decay
rates and sound periods for varying wavelength $\lambda_n= 2\pi/k_n$ are
compared with the theoretical relations in Fig.\ref{figcor2}.  Note that the
theoretical trend agrees quite well with the simulation results for $\lambda >
100$ \AA.  Considering that the size of one cell in these simulations is $15$
\AA, this means that one needs more that about 7 cells to properly resolve one
wave. In fact, for wavelengths $\lambda <100$ {\AA} the viscosity is
underestimated due to the reduction in spatial resolution. The same reasoning
applies to the sound time.

\section{Non-equilibrium states}
\label{non-eq}
In this section we present standard non equilibrium flow tests performed with
the mesoscopic Eulerian solver described in previous sections. In order to do
so, firstly we require the description of the explicit boundary conditions for
our fluid system enclosed between walls.
\subsection{Boundary conditions}
\label{BC}
\begin{figure}[tb!]
\begin{center}
  \includegraphics[width=7.5cm]{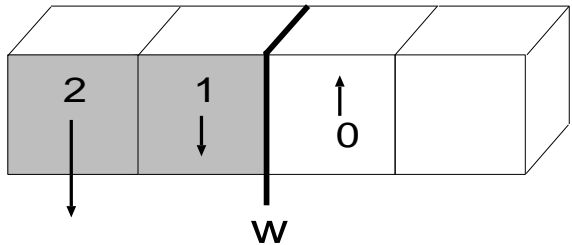}
\caption{Boundary conditions imposed at the interface ``$w$''. Cell 0
  represents a fluid cell, and cells 1 and 2 are {\em ghost
  cells}. Arrows indicate the velocity vectors in a case with zero
  velocity imposed at ``$w$'' (no-slip boundary condition and rigid wall
  at rest).}
\label{boundaryfig}
\end{center}
\end{figure}
The imposition of boundary conditions on the velocity is illustrated in
Fig.\ref{boundaryfig}. In the figure the boundary ``$w$'' is placed at the
interface between cells 0 and 1. The fluid region corresponds to cell 0 and
continues to the right, while cells 1 and 2 (in grey) are {\em ghost} cells
which are used to impose the desired mechanical behaviour at the boundary
``$w$''. In order to close the system, we need to evaluate the momentum flux
across the interface $w$. As at any other cell interface, we approximate
$\bm \Pi_w \cdot \bm e_{w}=[(\bm \Pi_0+\bm \Pi_1)/2]\cdot \bm e_{w}$, where
$\bm e_{w}$ is the surface unit vector (in this case $\bm e_{01}$). 
 Hence we require knowledge of $\bm
\Pi_1$, the stress tensor in the first {\em ghost} cell as  can also be 
inferred from Eq.  (\ref{presstens}) when evaluated at the boundary cell
``$0$''. According to the constitutive relation $\bm \Pi_1 \propto
\mbox{\boldmath$\nabla$} \bm v_1$.  Hence, a secondary {\em ghost} cell (\#$2$
in Fig.\ref{boundaryfig}) is required to evaluate the velocity gradient via
the central difference scheme: $\mbox{\boldmath$\nabla$} v_1\cdot \bm e_w =
(2\delta)^{-1} (\bm v_0 -\bm v_2)\cdot \bm e_w$ (where $\delta$ is the spatial
resolution).  This closes the set of
equations for the velocity at the {\em ghost} wall cells. This procedure
enables certain flexibility: one can either impose the value of the momentum
flux $\bm \Pi_w \cdot \bm e_{w}$ (von Neumann boundary condition), a
generalized relation involving the fluid velocity at the wall $\bm \Pi_w \cdot
\bm e_{w} \propto \bm v_w$ (Maxwell relation for fluid slip \cite{neto05}) or
the more standard no-slip condition, $\bm v_w = \bm U_{wall}$ (Dirichlet
boundary condition).  In the present work we assume no-slip at the wall and
set the velocity of the {\em ghost} cells accordingly to a linear interpolation
of the velocity; with $\bm v_w = \bm U_{wall}$ we find
\begin{eqnarray}
{\bm v}_1 &= 2\,{\bm U}_{wall} - {\bm v}_0,\nonumber\\
{\bm v}_2 &=  4\,{\bm U}_{wall} - 3 {\bm v}_0.
\end{eqnarray}
More general slip boundary conditions can be obtained by choosing a different value for
$\bm v_w$.  As is customary, the density at the wall is uniquely controlled by
the fluid, meaning that $M_w=M_0=M_1=M_2$ \cite{garcia87}.

\subsection{Couette, Poiseuille and cavity flows}
\label{couette_poiseuille_cavity}
\begin{figure}[tb!]
\begin{center}
\includegraphics[width=6cm,angle=0]{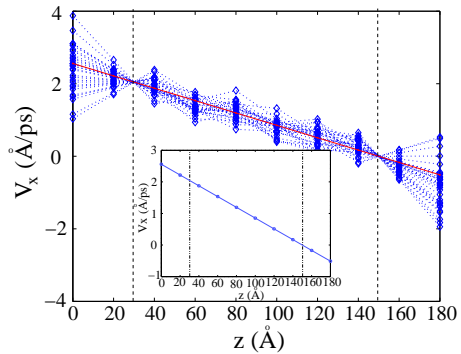}
\caption[]{Stationary Couette profile according to fluctuating hydrodynamics: 
  The diamonds are simulation results and the continuous line is the theoretical stationary
  linear profile. The wall velocity has been set $2.04$ \AA/ps.  The
  inset figure shows a deterministic Navier-Stokes simulation with the same
  parameters.  Vertical dashed lines represent the boundary walls.  }
\label{couette}
\end{center}
\end{figure}

\begin{figure}[tb!]
\begin{center}
\includegraphics[width=6cm,angle=0]{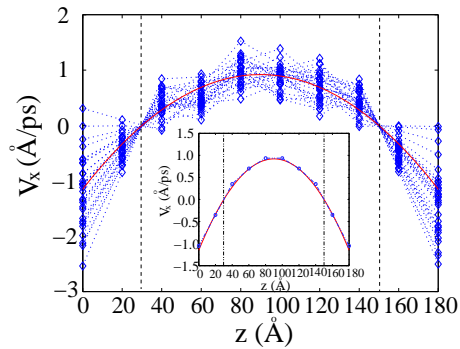}
\caption[]{Stationary Poiseuille profile according to fluctuating
  hydrodynamics: Diamonds correspond to simulation results and the continuous
  line is the theoretical profile associated with Navier-Stokes flow. The inset figure shows the same flow case
  for a purely deterministic simulation. In both figures, the vertical dashed
  lines represent the walls. }
\label{poiseuille}
\end{center}
\end{figure}

Stochastic and deterministic simulations for three different flow situations
(Couette, Poiseuille and cavity flow) have also been performed, displaying
good comparisons with theoretical predictions.
Figures \ref{couette} and \ref{poiseuille} show stationary fluctuating (and
deterministic) flows for argon at ambient temperature and mass density
$\rho_0=0.599$ g/mol/\AA$^3$. The simulations are performed using $10 \times 10
\times 10$ cells representing a periodic box of size $200 \times 200 \times
200$ \AA$^{3}$.  The fluid is confined in a channel defined by two infinite
parallel planes orthogonal to the $z$ axis. In this particular geometry the
first two and last two layers of particles in the $z$ direction are {\em ghost}
particles belonging to the walls; the no-slip condition is satisfied at
$z_1=30$ {\AA} and $z_2=150$ \AA, so the fluid is confined in a region of
width $120$ \AA.

A Couette flow is shown in Fig.\ref{couette}.  We plot the $x$ component of
the stationary velocity field in the $z$ direction.  The wall amplitude
velocity has been set at $2.04$ \AA/ps, while the amplitude of fluid velocity
fluctuations is about $0.5 $ \AA/ps, for the temperature and cell volume
considered.  The inset picture corresponds to an equivalent simulation but
with the thermal fluctuations  within the pressure tensor switched off. In this
limit, we recover standard Navier-Stokes behaviour.

Figure \ref{poiseuille} shows the $x$ component of the stationary velocity
field in the $z$ direction for a Poiseuille flow. The applied gravity force in
this case is $0.0174$ {\AA}/ps$^2$. The same simulation performed without
thermal fluctuations is displayed in the inset picture and both are  compared
with the theoretical solution.

\begin{figure}[tb!]
\begin{center}
\includegraphics[width=6cm]{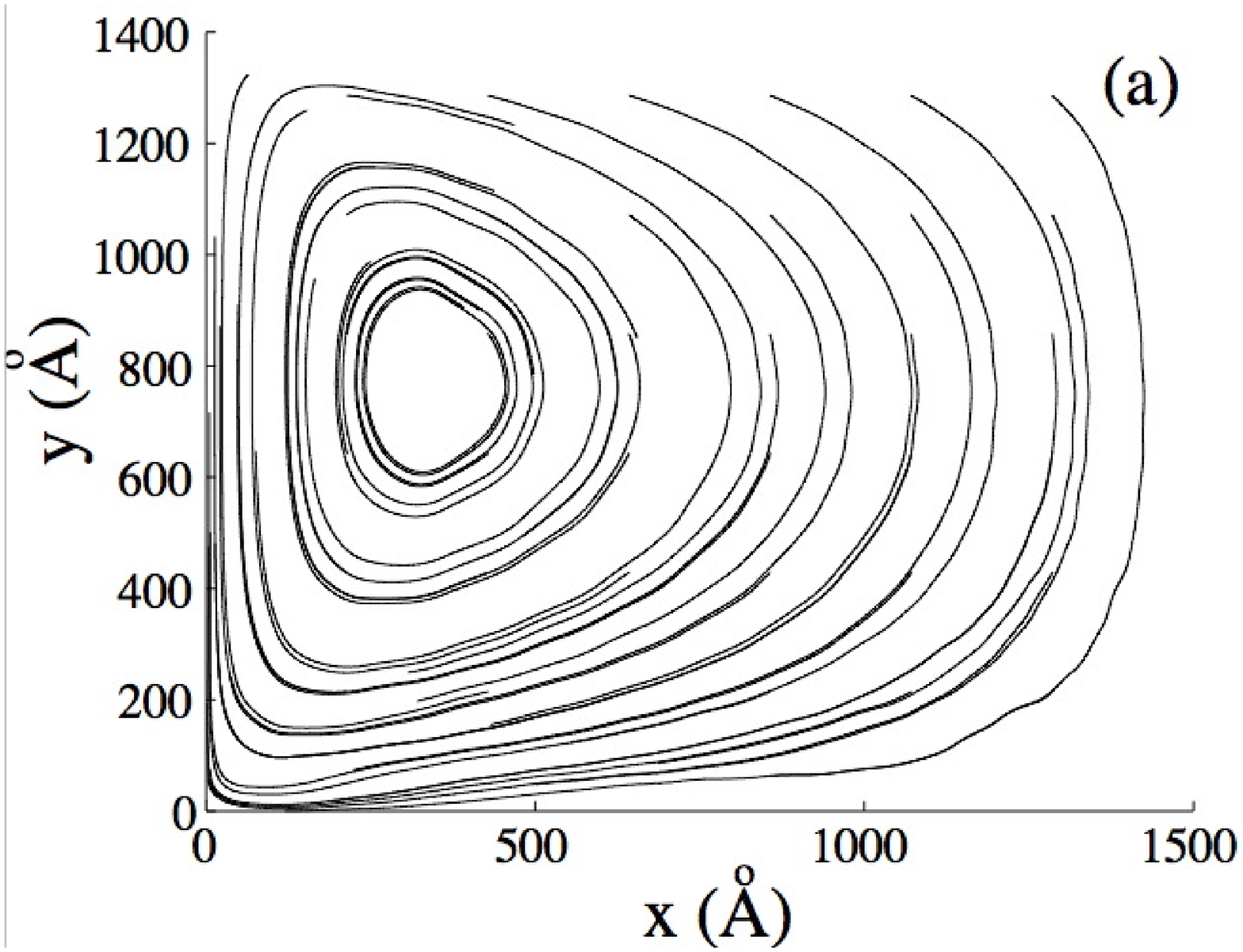}\\
\includegraphics[width=6cm]{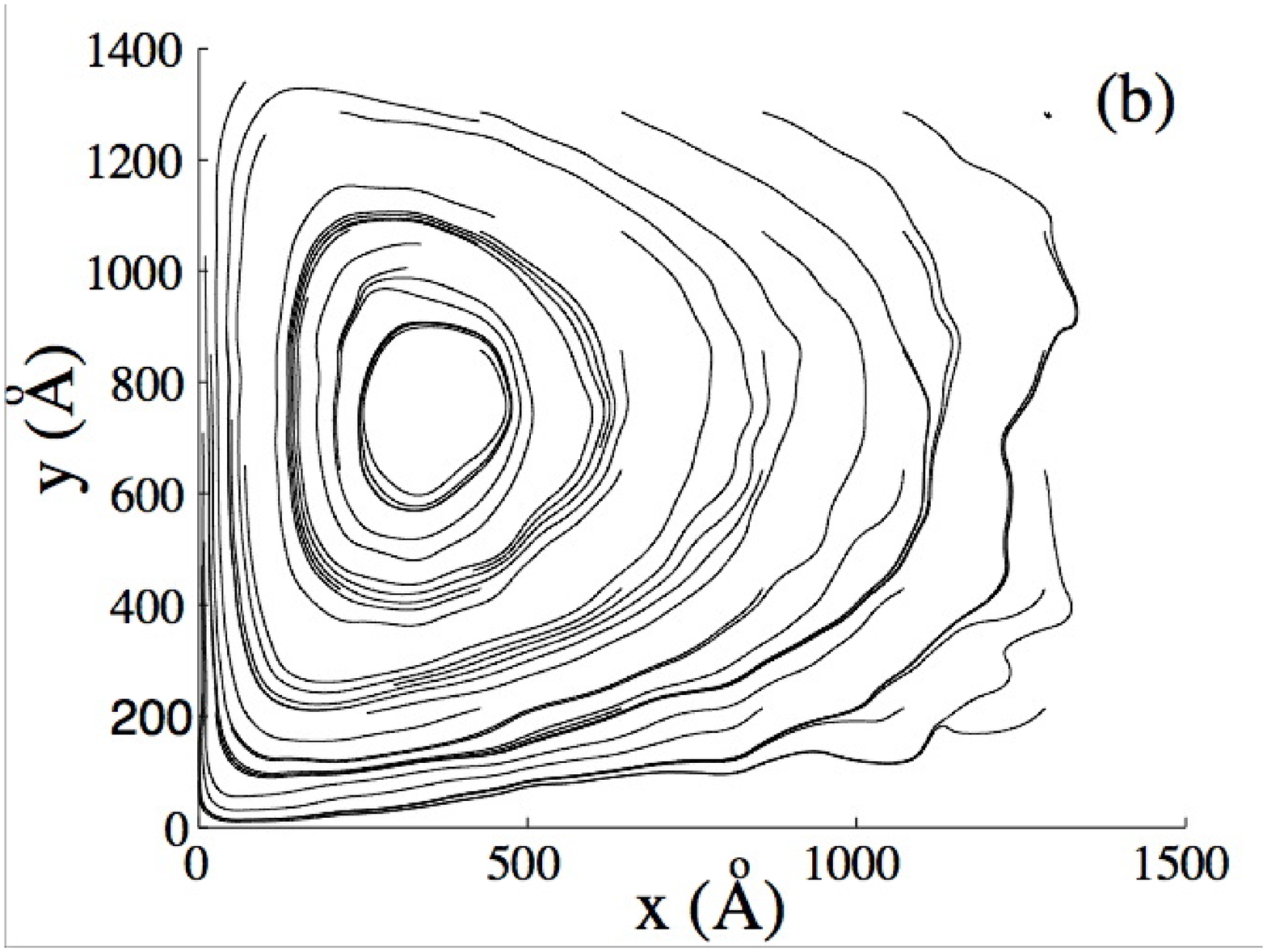}
\caption{Streamline plots of a cavity flow for TIP3P water model in the 
  stationary regime.  The wall velocity along $y$ is $1$ \AA/ps, and the
  temperature is $300$ K. The dimensions of the cavity are
  $1500\times1500\times50$ {\AA}$^3$ and the mesh  is $30\times30\times1$.
  In (a) thermal fluctuations of the pressure tensors are not present,
  and the Navier-Stokes solution recovered. In (b) we show the fluctuating hydrodynamics
  solution.  }
\label{cavityfig}
\end{center}
\end{figure}

We have also carried out simulations of cavity flow for (TIP3P) water depicted
in Fig.\ref{cavityfig} within a domain of dimensions $1500\times1500\times50$
{\AA}$^3$ and mesh  $30\times30\times1$. The wall is moving at a constant
speed of $1$ \AA/ps in the $y$ direction.
The average flow corresponding to averaging 
the fluctuating hydrodynamics result (Fig. \ref{cavityfig}b) corresponds to the 
fluctuation-less flow of Fig. \ref{cavityfig}a.
Although this is quite a large fluid domain, thermodynamic fluctuations are
still very visible and have a major effect on the flow.  For smaller domains
the stationary circulatory flow can be completely nullified by the
fluctuations.  This kind of cavities can be used as mixers in microfluidics
applications, but the extent of the mixing is  affected by the
magnitude of the fluctuations.  

\section{Summary}
\label{summary}
We have derived a finite volume discretization of the equations of fluctuating
hydrodynamics.  The model provides a good representation of the thermodynamic
fluctuations which are important at the mesoscale, typically in the nanometer
range, as well as the hydrodynamics.  We have tested the model in equilibrium
and non-equilibrium situations. Simple no-slip boundary conditions have been
used and tested in three flow situations Couette, Poiseuille and cavity
flow.  From a set of molecular dynamics simulations, we have also derived a
simple approximation to the equations of state for the TIP3P water model and
for argon which permit us to simulate these compressible fluids at ambient
temperatures ({\em ca} 300 K) and pressures around 1 atm. 

There are some aspects of the model which could be improved and we reserve
 for future work.  The discretization of the second order derivatives is
based on a central difference method.  This approach is  known to produce
instabilities which lead to a characteristic and undesirable alternating pattern
in the velocity and density fields \cite{patankar80}. However, we have found that such
problems only arise with very strong perturbations well beyond the typical flows at 
these scales. 
A further important extension is to incorporate the energy equation
so as to be able to model thermal phenomena.  Finally, we have used a very
simple boundary condition appropriate for the  cases studied here: more
sophisticated boundary conditions, which may reproduce diverse molecular
boundaries (e.g. taking into account the hydrophobic hydrophilic nature of
the walls) while also retaining the fluctuations, can be devised in the framework of 
the hybrid models \cite{defabritiis06hmd}.

Our mesoscopic fluctuating model serves to support a wide range of
applications. We are currently using it in hybrid molecular-continuum
simulations \cite{defabritiis06hmd}, with implicit solvent models
\cite{defabritiis06his}, and we plan to use it for the study of microfluidic
flows. The use of a regular lattice and the Eulerian description greatly simplify 
the implementation of this model in a serial (as here) or parallel computing environment. 
As a result, we  believe that it furnishes a unique tool to explore  hydrodynamics
 at the nanoscale including the effects of  fluctuations, in stand-alone more or 
coupled with molecular dynamics. 

{\bf Acknowledgements}

GDF and PVC are grateful to EPSRC (UK) for funding the Integrative Biology project
(GR/S72023).  MS is supported by the Spanish Ministerio de Educaci\'on y
Ciencia project FIS2004-01934 and by Programa Propio de Investigaci\'on de la
UNED (2006).  PVC \& MS thank  EPSRC for funding RealityGrid under grant
number GR/R67699, which supported MS's 6 month visit to the CCS at UCL during
2005.  RDB acknowledges support from the EU project MERG-CT-2004-006316 and
Spanish project CTQ2004-05706/BQU. We are indebted to A. Dejoan, P. Espa\~{n}ol,
E.  Flekkoy and S. Succi for helpful discussions.

\appendix
\section{Hydrodynamic modes}
\label{appendix1}
In the isothermal situations the equations that describe
the transport of mass and momentum density fields are
\begin{eqnarray}
\partial_t \rho&=& -\mbox{\boldmath$\nabla$}\cdot {\bf g},\nonumber\\
\partial_t {\bf g}&=& -\mbox{\boldmath$\nabla$}\cdot 
\left({\bf g}  {\bf v} \right)-\mbox{\boldmath$\nabla$}p
+\eta \mbox{\boldmath$\nabla$}^2 {\bf v}
+\left(\frac{\eta}{3}+\zeta\right) \mbox{\boldmath$\nabla$}
\left(\mbox{\boldmath$\nabla$} \cdot {\bf v}\right).\nonumber\\
\label{NS}
\end{eqnarray}
The equilibrium state is characterized by a constant density field $\rho_e$
and a zero momentum density field ${\bf g}_e=0$ because the fluid is at rest.
For fluctuations of small enough amplitude, the relaxation towards equilibrium
is governed by the linearized version of the mass and momentum Eqs.(\ref{NS}).
By decomposing the hydrodynamic fields as $\rho({\bf r},t)=\rho_e+\delta
\rho({\bf r},t)$ and ${\bf g}({\bf r},t)=\delta {\bf g}({\bf r},t)$, the
linearized version of Eqs.(\ref{NS}) for the perturbations results
\begin{eqnarray}
\partial_t \delta\rho&=& -\mbox{\boldmath$\nabla$}\cdot \delta{\bf g},\nonumber\\
\partial_t \delta{\bf g}&=& 
-\mbox{\boldmath$\nabla$}p+\nu \mbox{\boldmath$\nabla$}^2 \delta{\bf g}
+\nu_B \mbox{\boldmath$\nabla$}\left(\mbox{\boldmath$\nabla$} \cdot
\delta{\bf g}\right),
\label{NS2}
\end{eqnarray}
with the usual definitions for the kinematic shear viscosity
$\nu\equiv\eta/\rho_e$ and the effective bulk viscosity $\nu_B \equiv
(\eta/3+\zeta)/\rho_e$.

Under isothermal conditions the thermodynamic relation of pressure
perturbation $\delta p$ with density and temperature becomes simply
\cite{boon_yip},
\begin{equation}
\delta p(\rho,T)= c_T^2(\rho) \delta \rho,
\end{equation}
where $c_T^2 \equiv (\partial p/\partial \rho)_T$ is the squared
isothermal sound velocity.

Let us consider a general solution as a series of normal modes
\begin{equation}
\label{normalmode} \bm a({\bf r},t)=\bm a({\bm k},t) e^{i {\bm k}\cdot {\bf r}},\nonumber\\
\end{equation}
where we have gathered the
hydrodynamic variables in the array $\bm a = \left\{\delta \rho, \delta {\bf
    g}\right\}$.  Taking the Fourier transform in Eqs. (\ref{NS2}) one gets
the equations $\bm a({\bm k},t)$. In the linear regime there is no coupling
between modes and without loss of generality one can work in the reference
system for which the wave vector is ${\bm k}=(k,0,0)$,
\begin{equation}
\label{determ}
\frac{d{\bm a(k,t)}}{dt} = \bm{H} \bm a(k,t)
\end{equation}
where $\bm a(k,t) = \left(\rho(k,t), g_{x}(k,t), g_{y}(k,t), g_{z}(k,t)\right)^T$ and
the hydrodynamic matrix is
\begin{equation}
\label{hmatrix}
{\bm H} \equiv -
\left[
\begin{array}{cccc}
0 &i k&0 &0 \\
i c_T^2 k &\nu_L k^2&0 &0 \\
0 &0 & \nu k^2 &0 \\
0 &0 &0& \nu k^2  \\
\end{array}
\right],
\end{equation}
with the kinematic longitudinal viscosity defined as $\nu_L=\nu+\nu_B$.  The
eigenvalues of the hydrodynamic matrix $\bm H$ provide the growth rates of the
normal modes of the system given by Eq. (\ref{normalmode}).  The
eigenvalues are obtained from the roots of the characteristic equation
$\det[\bm H - \omega \bm 1] =0$, which results in
\begin{equation}
\left(\omega+\nu k^2\right)^2\left(\omega^2 +\nu_Lk^2\omega+c_T^2k^2
\right)=0.
\end{equation}
The solutions are
\begin{eqnarray}
\omega_{1,2}&=&-\nu k^2,\nonumber\\
\omega_{3,4}&=&-\Gamma_T k^2\pm i s_Tk,
\label{om34}
\end{eqnarray}
where we have defined $\Gamma_T$ as the isothermal sound absorption
coefficient and $s_T$ as the sound speed depending on the wave vector
given by
\begin{eqnarray}
\label{abs}\Gamma_T&=&\frac{\nu_L}{2},\nonumber\\
\label{st} s_T&=&\frac{\sqrt{4  c_T^2-\nu_L^2 k^2}}{2}.
\end{eqnarray}
The first two eigenvalues ($\omega_{1,2}$) correspond to the two shear modes
associated with the exponential decay of the transversal momentum $g_y$ and
$g_z$.  Sound modes correspond to $\omega_{3,4}$. Indeed, as can be seen from
Eqs. (\ref{om34}) and Eqs.  (\ref{st}), sound is underdamped if $s_T$ is a
real number. However, according to Eq. (\ref{st}) if $k>2c_T/\nu_L$, sound
becomes overdamped. Nevertheless for most liquids $c_T/\nu_L \sim O(1)$ so
this anomalous solution occurs at quite small wavelengths for which the
present mesoscopic description does not apply (at molecular lengthscales one
should consider the dependence of the transport coefficients on $k$ within the
generalized hydrodynamic formalism \cite{boon_yip}).  As a matter of fact, the
difference between $s_T$ and $c_T$ is negligible for any mesoscopic
wavelength, so throughout the present paper we assume that $s_T=c_T$.

With this last approximation, the solution is given by
\begin{eqnarray}
\label{dens}
\rho({\bm k},t)&=&\rho_e+\rho({\bm k},0)\, \exp\{-\Gamma_T k^2 t\}\cos(c_Tkt)\nonumber\\ 
&-&\frac{i}{c_T}\exp\{-\Gamma_T k^2 t\}\sin(c_Tkt)\hat{{\bm k}}\cdot{\bf g(\bm k,0)},\nonumber\\ 
{\bf g}(\bm k,t)&=&\exp\{-\Gamma_T k^2 t\}  
[\cos(c_Tkt) \hat{{\bm k}}\cdot{\bf g}(\bm k,0) \nonumber\\
&-&  i \sin(c_Tkt)\, c_T \rho({\bm k},0) ] \hat{{\bm k}}  \nonumber\\
&+&\exp\{-\nu k^2 t\}\left({\bf 1}-\hat{{\bm k}}\hat{{\bm k}}\right)\cdot {\bf g}(\bm k, 0)
\end{eqnarray}
where $\hat{{\bm k}}=\bm k/|\bm k|$ is the unit wave vector.

\section{Equations of state for argon and water via molecular dynamics simulation}
\label{Argon_water}
In this section we study the equations of state for water and also argon
through molecular dynamic simulations using the NAMD molecular dynamics
code \cite{NAMD}.  In particular, the theoretical Lennard-Jones equation of
state for argon given in Refs.\cite{Johnson93, Heyes88, Borgelt90} is not
necessarily exact because our MD simulations using the CHARMM force field
perturb the Lennard-Jones potential close to the cutoff radius, smoothing it to
zero \cite{NAMD}.  Thus we decided to obtain an accurate approximation of the
 equation of state by directly fitting the data obtained from molecular
dynamics simulations to a second order polynomial.  These considerations also
apply for the TIP3P water model.
\begin{figure}[tb]
\begin{center}
\includegraphics[width=7cm,angle=-90]{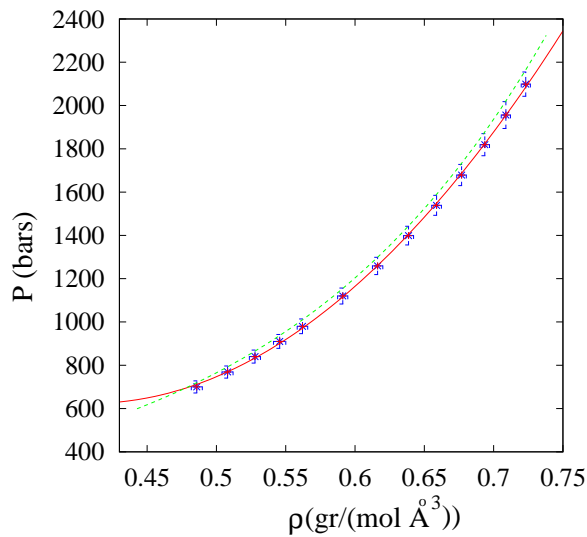}
\caption[]{Argon equation of state: Pressure (in bars) versus mass 
  density at temperature $300$K.  The simulation results appear with error
  bars and the dashed line is the equation of state for the theoretical
  Lennard-Jones fluid in Ref.\cite{Johnson93, Heyes88, Borgelt90}. The
  continuous line is the fit of the numerical data to the second order
  polynomial $3088.21-12065.2\rho+14765.8 \rho^2$.}
\label{press_argon_bar}
\end{center}
\end{figure}

\begin{figure}[tb]
\begin{center}
\includegraphics[width=7cm,angle=-90]{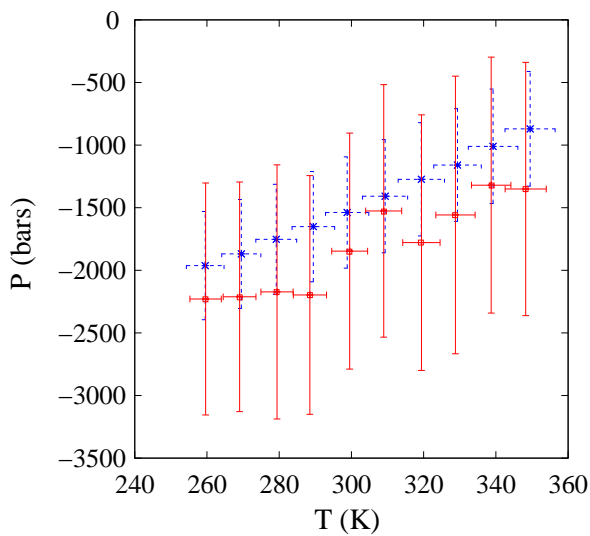}
\caption[]{Water pressure (in bars) versus temperature 
obtained from MD simulations of the TIP3P water model (using NAMD)
at a fixed density $\rho=0.55066$ g/(mol {\AA}$^3$). The $\ast$  
  symbols correspond to simulations with all bonds rigid while the $\Box$
  symbols are for non rigid bonds. 
}
\label{water_press_temp}
\end{center}
\end{figure}

We have computed the equation of state for argon in our molecular model in the
NpT ensemble at a fixed temperature $300 \pm 4$ K with $5000$ argon atoms. The
simulations are performed with a time step $\Delta t=1$ fs.  We have made use
of a switched Lennard-Jones potential \cite{NAMD} between 10 and 12 {\AA}.
The results are displayed in Fig.\ref{press_argon_bar} for the mass density.
Note that the deviations between the theoretical model and the simulation
results for the pressure can be as large as $15\%$ in the units presented in
the graph, the simulated argon pressure always being less than that given by
the theoretical Lennard-Jones equation of state in Ref.\cite{Johnson93,
  Heyes88, Borgelt90}.

Concerning the water equation of state, we have performed  simulations of
water molecules using the TIP3P water model \cite{Jorgensen}.  In
Fig.\ref{water_press_temp} we present the simulation results for a range of 
temperatures around $300$K with a time step $\Delta t=1$ fs.  We have chosen
a fixed density $\rho=0.55066$ g/(mol {\AA}$^3$) what gives (in a NVT ensemble
for a cubic periodic box of size $30$ \AA) a total of  826 water
molecules.  We have also tested the effect of using rigid bonds.  We see that
for  harmonic bonds the fluctuations in pressure are bigger compared to those of
the rigid simulation providing also smaller mean average pressures in general
terms.  We have also compared our results with three
analytical models for the water equation of state presented in Ref.
\cite{Jeffrey}. We observe clear deviations from the models.  Basically,
all three theoretical models largely overestimate the isothermal sound
velocity at any value of the density considered.  This reinforces the
necessity of pre-calibrating the equation of state for each particular fluid
considered; for instance via MD simulations as done here.

\begin{figure}[tb]
\begin{center}
  \includegraphics[width=7cm,angle=-90]{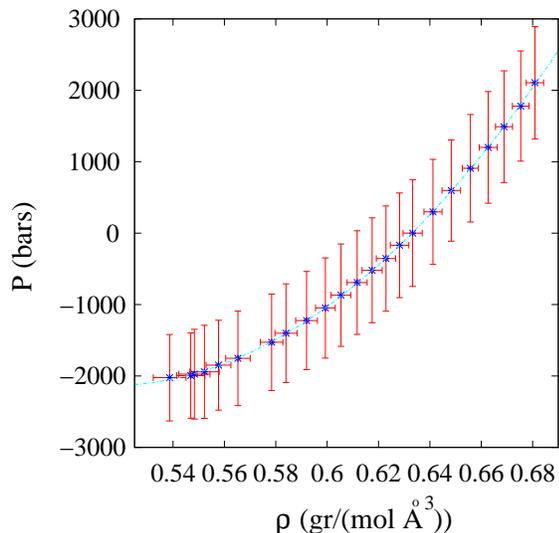}
\caption[]{Water pressure (in bars) versus mass density for a
  temperature of $300$ K from MD simulations of the non-rigid TIP3P
  water model (results obtained with the NAMD code).
  The continuous line is the best fit of the numerical data to the second order
  polynomial $p(\rho)=38373.6-157398\rho+152881\rho^2$ bars. 
}
\label{water_press_density}
\end{center}
\end{figure}

In Fig.\ref{water_press_density} we plot the values of mass densities obtained
from water MD simulations in an NpT ensemble against pressure (in bars) at a fixed
temperature $300 \pm 5$ K with non-rigid water molecules and a time step
$\Delta t=1$ fs.
The second order polynomial fit used in the FH equations is also shown
 in the continuous line.

\bibliographystyle{apsrev}
\bibliography{../../gianni}

\end{document}